\documentclass[runningheads]{llncs}
\usepackage{rotating}
\usepackage{array}
\usepackage{colortbl}
\usepackage{booktabs}
\usepackage{tablefootnote}

\usepackage[flushleft]{threeparttable}

\usepackage{comment}
\usepackage{amsfonts}
\usepackage{graphicx}
\usepackage{textcomp}
\usepackage{booktabs}
\usepackage{tabularx}
\usepackage{booktabs}
\usepackage{multirow} 
\usepackage{graphicx}
\usepackage{float}
\usepackage{lipsum}
\usepackage{msc}
\usepackage{xcolor}
\usepackage{tcolorbox} 
\usepackage{soul} 
\usepackage{enumitem}
\usepackage{pifont}
\usepackage{graphicx}
\usepackage{caption}

\usepackage{wrapfig}
\usepackage{booktabs}

\usepackage{floatflt}

\tcbuselibrary{listingsutf8}

\newtcolorbox{roundedtextbox}[1][]{%
 colback=black!5!white,
 colframe=black!40!white,
 fonttitle=\bfseries,
 arc=10pt, 
 boxrule=1pt,
 fontupper=\small,
 #1
}

\usepackage{hyperref}
\hypersetup{colorlinks=true,linkcolor=blue,urlcolor=blue}

\usepackage[frozencache=true,cachedir=minted-cache]{minted} 

\definecolor{verylightgray}{rgb}{0.95, 0.95, 0.95}

\usepackage{circledsteps}

\usepackage{tikz}
\usepackage{xcolor}
\pgfkeys{/csteps/inner ysep=2pt}
\pgfkeys{/csteps/inner xsep=5pt}

\newcommand{\quicAflnetImpCount}{6}
\newcommand{\bugcountQUICAFLNet}{10}

\newcommand{\bugcountQAlfonly}{6} 

\newcommand{\sut}{{\textsf{QUT}}}

\usepackage[normalem]{ulem}

\definecolor{BackgroundBlue}{RGB}{218, 232, 252}
\definecolor{BackgroundYellow}{RGB}{255, 242, 204}
\definecolor{BackgroundOrange}{RGB}{255, 230, 204}
\definecolor{BackgroundRed}{RGB}{248, 206, 204}
\definecolor{BackgroundGreen}{RGB}{213, 232, 212}
\definecolor{BackgroundPurple}{RGB}{225, 213, 231}

\usepackage[T1]{fontenc}
\newcommand{\pmsg}[1]{{\fontfamily{lmss}\selectfont{#1}}}

\newcommand{\tgt}[1]{{\fontfamily{cmss}\selectfont{#1}}}

\newcommand{\nameQUICAFLNet}{\textsc{QUIC-Fuzz}}
\newcommand{\fn}{\textsc{Fuzztruction-Net}}
\newcommand{\aflnet}{\textsc{ALFNet}}
\newcommand{\chatafl}{\textsc{ChatAFL}}

\usepackage{xcolor}
\usepackage{tcolorbox}
\usepackage[framemethod=TikZ]{mdframed}

\usepackage{tikzpeople}

\usepackage[T1]{fontenc}
\usepackage{graphicx}
\begin{document}
\title{\nameQUICAFLNet: An Effective Greybox Fuzzer For The QUIC Protocol}
%
%
%
\author{Kian Kai Ang \and
Damith C. Ranasinghe
}
\authorrunning{K. K. Ang and D. C. Ranasinghe}
%

\institute{University of Adelaide, Australia
\email{\{kiankai.ang, damith.ranasinghe\}@adelaide.edu.au} 
}

\maketitle              
\begin{abstract}
Network applications are routinely under attack. 
We consider the problem of developing an effective and efficient fuzzer for the recently ratified QUIC network protocol to uncover security vulnerabilities.
QUIC offers a \textit{unified} transport layer for low latency, reliable transport \textit{streams} that is \textit{inherently} secure, ultimately representing a complex protocol design characterised by new features and capabilities for the Internet. Fuzzing a \textit{secure transport layer} protocol is not trivial. The interactive, strict, rule-based, asynchronous nature of communications with a target, the stateful nature of interactions, security mechanisms to protect communications (such as integrity checks and encryption), and inherent overheads (such as target initialisation) challenge \textit{generic} network protocol fuzzers. We discuss and address the challenges pertinent to fuzzing transport layer protocols (like QUIC), developing mechanisms that enable fast, effective fuzz testing of QUIC implementations to build a prototype \textit{grey-box} mutation-based fuzzer---\nameQUICAFLNet{}. We test \quicAflnetImpCount{}, \textit{well-maintained} server-side implementations, including from Google and Alibaba with \nameQUICAFLNet. The results demonstrate the fuzzer is both highly effective and generalisable. Our testing uncovered \textbf{\textit{\bugcountQUICAFLNet{} new}} security vulnerabilities, precipitating 2 CVE assignments thus far.  In code coverage, \nameQUICAFLNet{} outperforms other existing state-of-the-art network protocol fuzzers---\fn{}, \chatafl{}, and \aflnet{}---with up to an 84\% increase in code coverage where \nameQUICAFLNet{} outperformed \textit{statistically significantly} across all targets and with \textit{a majority of bugs} only discoverable by \nameQUICAFLNet{}. We open-source \nameQUICAFLNet{} on GitHub \textsf{\href{https://github.com/QUICTester/QUIC-Fuzz}{https://github.com/QUICTester/QUIC-Fuzz}}. 
\end{abstract}
\section{Introduction} \label{sec:quic_aflnet_introduction}
\vspace{-1mm}
QUIC is a \textit{secure} transport layer protocol \textit{optimised} for performance. It currently provides transport layer services for HTTP/3. QUIC establishes \textit{inherently} secure communication channels, ensuring message confidentiality, integrity, and availability for Internet applications over UDP while also providing a reliable transport layer service. Effectively, the protocol design aims to reduce the latency and connection overheads associated with the use of TLS~\cite{RN49} over TCP for secure transport~\cite{RN43}. However, the services provided by QUIC differ from TCP and TLS over TCP and include the ability to have data available before a secure handshake is completed, the possibility of multiple simultaneous streams, and the provision of five different security levels for connections. Importantly, network applications provide external facing interfaces to the world, making them a common attack vector for adversaries to exploit. Given QUIC's role in various Internet applications---driven largely by its use in HTTP/3---implementation bugs within QUIC can severely affect the security and quality of Internet applications. 

Despite research to advance fuzzing and its proven effectiveness at uncovering vulnerabilities by automatically generating and injecting inputs to test software systems, adopting fuzzing techniques to test network protocols is not straightforward. Recent efforts have led to methods to address inherent challenges facing fuzz testing network protocols~\cite{aflnet,SGF,natella2022stateafl}, including the investigation of LLMs to automate fuzzing processes, such as generating test sequences~\cite{chatafl}. However, many of these techniques and tools have been developed to test application-layer protocols, leaving transport layer protocols (such as QUIC), which provide reliable and secure services, relatively under-explored. This is because testing secure transport protocols poses unique challenges.

\vspace{2px}
\noindent\textbf{Challenges.}
\pgfkeys{/csteps/inner color=white}
\pgfkeys{/csteps/fill color=black}
\vspace{-4px}
\begin{description}
  \item [\Circled{C1}]\textbf{Input Integrity and Interpretability.~}First, secure services in QUIC are built with features to support five different security levels.  Cryptographic operations and checksums support security services, such as authentication and message confidentiality. Consequently, a fuzzer must account for the negotiated encryption secrets and cryptographic techniques when generating mutated inputs to exercise deeper functionality, code, and states. Current state-of-the-art replay-based fuzzers~\cite{aflnet,SGF,qin2023nsfuzz} apply direct mutations to recorded messages, corrupting encrypted messages and compromising message integrity. Such inputs fail integrity checks and are not meaningfully interpreted at a target; instead, are simply discarded without further processing. This impacts the \textit{effectiveness} of the fuzzer.

  \item [\Circled{C2}]\textbf{Non-determinism.~}Second, QUIC is \textit{complex} and \textit{unique}. It provides secure and reliable services with a holistic protocol. QUIC implementations optimise performance to manage reliable and secure service paths while servicing multiple clients setting up multiple streams with endpoints---and fuzzing such a protocol introduces non-deterministic behaviours at a target that can hinder making incremental fuzzing progress. For instance, when a target receives the same input at different time intervals influenced by factors---such as the resource scheduling and timeouts used in reliable data transfer---the target may exercise a different code path. Therefore, fuzzers may not be able to reliably use the same test case to reach a previously observed, interesting state that would allow mutation-guided input generation methods to make incremental progress. This can impact the \textit{effectiveness} of the fuzzer by impeding its ability to explore deeper code and new protocol states.

  \item [\Circled{C3}]\textbf{Test Execution Overhead.~}Third, in fuzz testing, the need to spawn a new target instance for every fuzz test introduces a significant initialisation overhead when dealing with network protocol implementations. These overheads can include loading the shared libraries, loading and phrasing configuration files, and initialising cryptographic primitives. The overhead significantly impacts testing \textit{efficiency}; the test case executions per second.
\end{description}

\noindent\textbf{Our Work.}
QUIC's growing prominence motivates us to address these challenges to build an \textit{effective} and \textit{efficient} fuzzer for testing the IETF-ratified QUIC implementations capable of uncovering software bugs and potential security vulnerabilities. To this end, we design and build \nameQUICAFLNet{}---a grey-box mutation-based fuzzer that: i)~performs mutations on recorded encrypted QUIC messages without breaking their integrity, ii)~enables fast synchronous communication with the target to address barriers to fuzzing progress from non-deterministic behaviours, and iii)~mitigates the target initialisation overhead.

In particular, we developed a QUIC-specific cryptographic module to strategically decrypt and re-encrypt messages within the fuzzer's workflow, allowing the fuzzer to modify the actual message content while preserving message integrity. To minimise non-deterministic behaviour at the target, we integrate a synchronisation protocol to serialise communication between the fuzzer and the target. Effectively ensuring every event triggered by the same test case occurs in the same order at the target. Then, we integrate a snapshot protocol to determine when to capture a pre-initialised target state, which can be restored at the start of each fuzzing iteration to remove the initialisation overhead. In this study, we focus on fuzz testing the more impactful server-side implementations of QUIC because the same protocol library is used by clients and servers while server failures affect multiple active connections, as in DoS attacks.

\noindent\textbf{Contributions.~} We summarise the contributions made in this work as follows:

\vspace{-2mm}
\begin{itemize}[itemsep=2pt,parsep=1pt,topsep=3pt,labelindent=0pt,leftmargin=5mm]
  \item We propose \nameQUICAFLNet{}, a grey-box fuzzer for QUIC. 
  
  \item To enable effective mutations on encrypted and integrity-protected QUIC messages, we develop a QUIC-specific cryptographic module. To increase fuzzing efficiency: i)~we integrate a synchronisation protocol to allow fast synchronous communication between the fuzzer and the target, then ii)~a snapshot protocol, to determine strategic locations to save server state to reduce target initialisation overheads. 
  
  \item We open-source \nameQUICAFLNet{} at {\small \textsf{\url{https://github.com/QUICTester/QUIC-Fuzz}}}. 
\end{itemize}

\noindent\textbf{Findings.~}We have used \nameQUICAFLNet{} to test \quicAflnetImpCount{} publicly available, well-maintained QUIC servers---including Google and Alibaba (as listed in Table~\ref{tab:quic-aflnet_server_tested})---to assess the fuzzer's effectiveness and help contribute to improving the security and robustness of QUIC implementations as well as the safety of end-users.
\begin{itemize}[itemsep=2pt,parsep=1pt,topsep=3pt,labelindent=0pt,leftmargin=5mm]
  \item Thus far, \nameQUICAFLNet{} has uncovered \bugcountQUICAFLNet{} bugs with a \textit{bug bounty award} and two CVE assignments.
  \item Importantly, the state-of-the-art baseline fuzzers discovered only four of the bugs found by \nameQUICAFLNet{} in ten executions over the course of 48-hour-long fuzzing campaigns.
\end{itemize}

\noindent\textbf{Responsible Disclosure.~}Following the practice of responsible disclosure, we shared our findings with corresponding development teams by sending bug reports to either vendors or developers in accordance with their reporting policies. We summarise the current state of disclosures and vendor responses in Table~\ref{tab:quic_aflnet_faults}. 

\section{Network Protocol Fuzzing Primer} \label{sec:quic-aflnet_primer}
\vspace{-2mm}
We provide a brief overview of the QUIC protocol handshake to understand the process of secure connection establishment (our focus) and recent optimisations for \aflnet{}, Snapfuzz, and then delve into our framework in Section~\ref{sec:quicAflNetDesign}.

\vspace{-4mm}
\subsection{QUIC}\label{sec:quicHandshake}

\vspace{-3mm}
An entity using QUIC must complete a handshake with its endpoints before it can communicate. Unlike TLS over TCP, where the negotiation of transport and cryptographic parameters is performed separately,
QUIC combines transport and cryptographic parameter negotiations into a single handshake using \textit{packets}, \textit{frames} and \textit{messages} defined in~\cite[Section 12.4]{9000} carrying different data types. To simplify our explanations, we refer to the frames and messages encapsulated within the frames---such as CRYPTO frames---as simply \textit{messages}.

\begin{wrapfigure}{l}{0.55\textwidth}
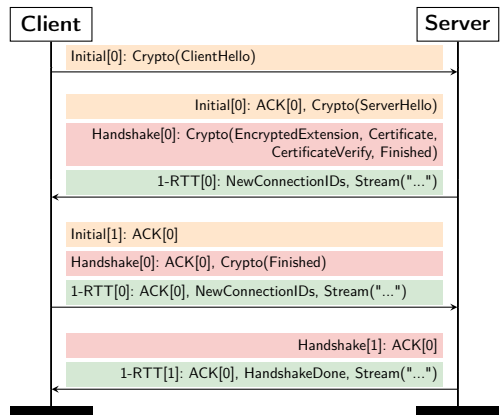

\vspace{-25pt}
\centering
 \resizebox{0.55\columnwidth}{!}{
    \begin{msc}[instance distance = 7.5cm, draw frame = none, environment distance = 0, instance width = 1.1cm, msc keyword=, head top distance=0cm]{} %
    \centering
    \declinst{client}{}{\parbox[c][1cm][c]{1.75cm}{
        \vspace{0.3cm}
        \centering
        \Large\textsf{\textbf{Client}}
    }}
    \declinst{server}{}{\parbox[c][1cm][c]{1.75cm}{
        \vspace{0.3cm}
        \centering
        \Large\textsf{\textbf{Server}}
    }}
    \nextlevel[0.25]
    \mess[label distance = 0.3ex, align=left]{\colorbox{BackgroundOrange}{\parbox{8.5cm}{\raggedright \pmsg{{Initial[0]: Crypto(ClientHello)}}}}}{client}{server}
    \nextlevel[5.8]
    \mess[label distance = 0.3ex, align=right, draw]{
    \colorbox{BackgroundOrange}{\parbox{8.5cm}{\raggedleft \pmsg{{Initial[0]: ACK[0], Crypto(ServerHello)}}}}\\[0.5ex]
    \colorbox{BackgroundRed}{\parbox{8.5cm}{\raggedleft \pmsg{{Handshake[0]: Crypto(EncryptedExtension, Certificate,\\
    CertificateVerify, Finished)}}}}\\[0.5ex]
    \colorbox{BackgroundGreen}{\parbox{8.5cm}{\raggedleft \pmsg{{1-RTT[0]: NewConnectionIDs, Stream("...")}}}}
    }{server}{client}
    \nextlevel[5.1]
    \mess[label distance = 0.3ex, align=left]{
    \colorbox{BackgroundOrange}{\parbox{8.5cm}{\raggedright \pmsg{{Initial[1]: ACK[0]}}}}\\[0.5ex]
    \colorbox{BackgroundRed}{\parbox{8.5cm}{\raggedright \pmsg{{Handshake[0]: ACK[0], Crypto(Finished)}}}}\\[0.5ex]
    \colorbox{BackgroundGreen}{\parbox{8.5cm}{\raggedright \pmsg{{1-RTT[0]: ACK[0], NewConnectionIDs, Stream("...")}}}}
    }{client}{server}
    \nextlevel[3.8]
    \mess[label distance = 0.3ex, align=right]{
    \colorbox{BackgroundRed}{\parbox{8.5cm}{\raggedleft \pmsg{{Handshake[1]: ACK[0]}}}}\\[0.5ex]
    \colorbox{BackgroundGreen}{\parbox{8.5cm}{\raggedleft \pmsg{{1-RTT[1]: ACK[0], HandshakeDone, Stream("...")}}}}
    }{server}{client}
    \end{msc}
    }
     \caption{QUIC \textit{Basic} handshake with \colorbox{BackgroundOrange}{Initial}, \colorbox{BackgroundRed}{Handshake} and \colorbox{BackgroundGreen}{1-RTT} packets. Packet numbers for 
     are inside the square brackets.}
    \label{2fig:quicBasicHandshake}
\vspace{-26pt}
\end{wrapfigure}

\vspace{-5mm}
\subsubsection{QUIC Handshake} QUIC provides \textbf{five} different security configurations, namely: i)~Basic (illustrated in Figure~\ref{2fig:quicBasicHandshake}); ii)~Client address validation without client authentication; iii)~Client authentication without address validation ; iv)~Client address validation and authentication; and v)~Handshake with a pre-shared key. We illustrate the Basic handshake in  Figure~\ref{2fig:quicBasicHandshake}. At the beginning, a client sends the server an Initial packet with a \pmsg{ClientHello} message containing application protocol negotiation, transport parameters, and the cryptographic information required to perform the key exchange. The server continues the handshake with an Initial packet and a Handshake packet. The Initial packet contains a \pmsg{ServerHello} message containing the cryptographic information needed to complete the secret exchange. The Handshake packet carries an \pmsg{EncryptedExtensions} message that contains transport parameters and the negotiated application protocol version, a \pmsg{Certificate} message that contains the server's certificate, a \pmsg{CertificateVerify} message that is used to request that the client verify the server's certificate, and a \pmsg{Finished} message that carries \pmsg{verify\_data}, a keyed hash computed using the handshake secret combined with the \textit{transcript hash} of all the previous TLS messages exchanged. For example, in Figure~\ref{2fig:quicBasicHandshake}, the server uses the \textit{transcript hash} comprising \pmsg{ClientHello}, \pmsg{ServerHello}, \pmsg{EncryptedExtensions}, the server's \pmsg{Certificate}, and the server's \pmsg{CertificateVerify} messages.

Once a server sends the \pmsg{Finished} message, it can transmit application data using the Stream frame in 1-RTT packets. The client continues the handshake by verifying the server's certificate and the \pmsg{verify\_data} in the server's \pmsg{Finished} message for authentication, key confirmation, and handshake integrity. The client then computes its \pmsg{verify\_data} using the handshake secret combined with the \textit{transcript hash} that includes all TLS messages exchanged until the server's \pmsg{Finished} message. 
The client includes the \pmsg{verify\_data} in its \pmsg{Finished} message and sends it to the server. Similarly, the server verifies the \pmsg{verify\_data} in the client's \pmsg{Finished} message for authentication, key confirmation, and handshake integrity. Following the successful verification of the \pmsg{verify\_data}, the server sends the client a 1-RTT packet with a \pmsg{HandshakeDone} message to indicate that the handshake is confirmed and the QUIC connection established. Notably, in contrast to TLS over TCP relying on IP addresses and port numbers, a QUIC connection is identified by a set of \pmsg{connection ID}s. 
Further \pmsg{connection ID}s for use post-handshake are exchanged through the \pmsg{NewConnectionID} messages.

\vspace{-4mm}
\subsection{Snapfuzz for ALFNet}\label{sec:snapfuzz_background}
\vspace{-1mm}
Snapfuzz~\cite{snapFuzz} proposes an optimisation framework for \aflnet{}~\cite{aflnet}.  
\aflnet{} is a grey-box mutation-based fuzzer specifically designed for stateful network protocols (see Appendix~\ref{apd:aflnet}). Effectively, \aflnet{} operates as the client application and directs its generated inputs in a sequence through network sockets (e.g. TCP/IP, UDP/IP) to the target. The inputs are generated by mutating a seed (sequences of recorded message exchanges between a client and server). In contrast to traditional fuzzers, \aflnet{} introduces the concept of using the SUT's state feedback to guide the fuzzing process. It uses a modified form of edge coverage combined with state awareness to efficiently identify interesting inputs that alter the control flow of the target application to help explore new code and states. 

Snapfuzz framework features two important components:
i) transforming slow asynchronous network communications between a fuzzer and a target into fast synchronous communication to eliminate synchronisation delays and ii) automatically deferring the creation of the Forkserver (responsible for spawning a new target instance for fuzzing) until the latest possible safe point. Snapfuzz implements these ideas with binary re-writing techniques to intercept \texttt{read()} and \texttt{write()} type syscalls to redirect them to custom system call handlers.

We adopt the Snapfuzz-\aflnet{} framework to address the pertinent challenges associated with fuzzing QUIC. Importantly, we address the limitations of current frameworks to build an \textit{efficient} and \textit{effective} fuzzer for QUIC targets.

\section{\nameQUICAFLNet{} Design} \label{sec:quicAflNetDesign}

\pgfkeys{/csteps/inner color=white}
\pgfkeys{/csteps/fill color=black}
\begin{figure*}[b!]
  \centering
  \includegraphics[width=1.0\linewidth]{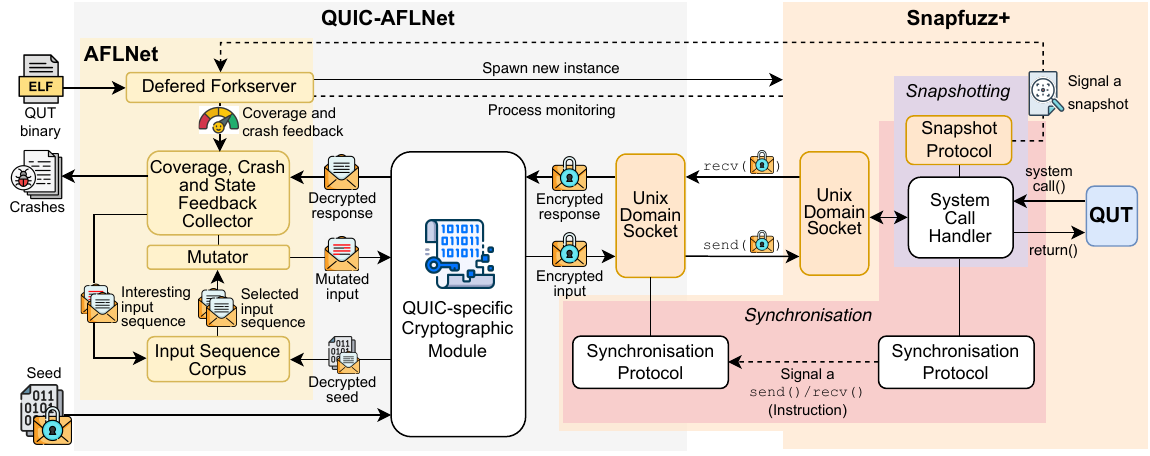}
  \caption{An overview of \nameQUICAFLNet{}. The QUIC-specific Cryptographic Module enables the fuzzer (in our prototype, ALFNet) to generate integrity-protected inputs. The Synchronisation Protocol allows synchronous communication between the fuzzer and the target (QUT) to address issues arising from non-determinism and delays impacting fuzzing throughput. The Snapshot Protocol determines a strategic location to save state to reduce server initialisation overhead.} 
  \label{fig:quic-AFLNet}
\end{figure*}

\vspace{-3mm}
Our mutation-based grey box fuzzer, \nameQUICAFLNet{}, integrates three key components to address the challenges discussed in Section~\ref{sec:quic_aflnet_introduction}:

\vspace{-2mm}
\begin{enumerate}
  \item QUIC-specific Cryptographic Module to generate integrity-protected inputs.
  \item Synchronisation Protocol to reduce delays and synchronise communication between the fuzzer and the target. 
  \item Snapshot Protocol to automate saving server state at strategic locations to reduce server initialisation overhead.
  \end{enumerate}

\vspace{-1mm}
In this section, we discuss how the design addresses fuzzing challenges \textbf{C1}-\textbf{C3}; Figure~\ref{fig:quic-AFLNet} presents an overview of \nameQUICAFLNet{} and the fuzzing loop.

\vspace{2mm}
\label{sec:quicCryptoModule}
\noindent\Circled{C1}\textbf{~Input Integrity and Interpretability.~}A Basic QUIC connection involves three different encryption levels for secure communication. At each encryption level, the client and server use the respective keys for encryption, encoding, and checksums. If either peer fails to validate an incoming protected message, the message is discarded without further processing.

The same principle applies to fuzzing: If the generated input cannot be recognised as an integrity-protected message, it is discarded before exercising interesting functionality. However, replay-based fuzzers (e.g. \aflnet{}) apply direct mutations to the seed (encrypted messages recorded during normal communication), do not alter actual content, and consequently invalidate the integrity of the generated input. In addition, fuzzers like \aflnet{} rely on the target's responses as feedback to guide state exploration. Because the responses from the \sut{} (QUIC Under Test) are also encrypted, the fuzzer cannot effectively extract detailed state information. The only information the fuzzer can collect and use as state feedback is the packet type, an unprotected field in the packet header. However, the packet type alone is insufficient for accurately inferring the \sut{}'s state. This is because packets of the same type can carry different content. For example, an Initial packet can carry either a PING, a CRYPTO, or a CONNECTION\_CLOSE frame. These roadblocks limit the effectiveness of a replay-based fuzzer, including inhibiting its ability to exercise deeper protocol codes and states.

\noindent\textbf{Methods for Resolving C1.~}We design a QUIC-specific Cryptographic Module to manage all cryptographic operations on the inputs and responses. Recall that the seed contains sequences of messages exchanged in a previous QUIC session. Before entering the fuzzing loop, the module decrypts the seed the user provides. The fuzzer then stores the decrypted input sequence as plain text in the Input Sequence Corpus. Because the input is now in plaintext, the fuzzer can safely mutate it without compromising the seed's integrity.

To ensure the \sut{} receives integrity-protected inputs, the module attempts to encrypt the input before sending it to the \sut{}. Notably, the encryption process may fail, particularly when the mutated packet structure is malformed. In such cases, the fuzzer opts to send the mutated input without encryption. This action is acceptable because unencrypted inputs are unlikely to be favoured in subsequent input scheduling if they do not contribute to uncovering new state or coverage. Consequently, it provides useful feedback to guide the fuzzer. In addition, the module decrypts all responses from the \sut{}, allowing the fuzzer to extract valuable state information to accurately infer the \sut{}'s state. 

However, to decrypt a seed, the cryptographic module requires the same secrets from the previous connection because the seed contains sequences of messages exchanged in a previous QUIC session. Further, the QUIC-specific Cryptographic Module and the \sut{} must use the same secret for encryption. If they do not share the same secret, the encrypted inputs sent by the fuzzer would be rejected by the \sut{} as invalid, and the fuzzer would be unable to decrypt and analyse the \sut{}'s responses. We employ the following methods to ensure that the cryptographic module and the \sut{} maintain the same secrets as those used during seed generation for Initial, Handshake, and 1-RTT encryption levels.

As mentioned in~\cite[Section 5.2]{9001}, any QUIC peer (including the \sut{}) can derive the Initial secret using \texttt{HKDF-Extract} with a default salt value and the Destination Connection ID (DCID) from the client's first Initial packet header. Therefore, we have designed the cryptographic module to perform the same operation as the \sut{}, extracting the DCID from the first packet in the seed to derive the Initial secret. But, unlike the Initial secret, which can be derived using information from the packet header, the Handshake and 1-RTT secrets are dynamically negotiated between the client and server during the handshake. This precludes the possibility of deriving these secrets directly from the seed. Therefore, we manually configure the Handshake and 1-RTT secrets used during seed generation into the QUIC-specific Cryptographic module and \sut{} to ensure that they consistently use the same secrets for cryptographic operations.

\vspace{1mm}
\label{sec:syncAndPerformance}
\noindent\Circled{C2}\textbf{~Non-determinism.~}As explained in Section~\ref{sec:quic_aflnet_introduction}, communications between the fuzzer and the \sut{} need to be synchronised to prevent non-deterministic events when replaying the same test case. For example, if the fuzzer acknowledges the \sut{}'s response too early while the response remains in flight or fails to acknowledge the \sut{}’s response in time, the \sut{} executes a different path to re-transmit the response. This is expected, especially in protocols that provide reliable data transfer, such as QUIC. 

A simple mitigation is setting a user-specified timeout to wait for the \sut{}’s response before sending the next input. But, the response may still arrive after the timeout expires. Further, such a solution sees each fuzzing iteration send a sequence of inputs and wait for a timeout after each input, with the delays significantly reducing the fuzzing throughput. In our ablation studies (see Section~\ref{sec:quic-aflent_stability}), we demonstrate the impact of eliminating these delays in \nameQUICAFLNet{}.

\vspace{1mm}
\noindent\textbf{Methods for Resolving C2.~}We adopt a Synchronisation Protocol to coordinate communication between \nameQUICAFLNet{} and the \sut{}. Specifically, the protocol monitors the send-and-receive operations on the \sut{} and signals appropriate instructions to the fuzzer. For example, when the \sut{} begins to receive data, the Synchronisation Protocol signals the fuzzer to send an input. Similarly, when the \sut{} performs a send operation, the protocol signals the fuzzer to receive the \sut{}'s response. Notably, applying the Synchronisation Protocol eliminates the need for manual timeout settings.

\vspace{1mm}
\noindent\Circled{C3}\textbf{~Test Execution Overhead.~}QUIC servers often require a long initialisation phase---we observed $\sim1$ to 30~ms periods across the \sut{}s we tested---before a target is ready to accept a connection. The initialisation process includes time for parsing configuration files, loading shared libraries, configuring security features, setting transport parameters, and binding to a network socket. Repeating this process in each iteration introduces significant overheads to fuzz testing a target and reduces the fuzzing throughput, that is, the fuzzer's \textit{efficiency}.

\vspace{1mm}
\noindent\textbf{Methods for Resolving C3.~}When initialising the Forkserver, we integrate a Snapshot Protocol to monitor the \sut{} initialisation process via its system calls. Effectively, we determine when the \sut{} is ready to accept a connection
and signal the Forkserver to capture a snapshot of the \sut{}. Subsequently, the Forkserver can spawn new \sut{} instances directly from the snapshot in each fuzzing iteration to remove the initialisation overhead.

\vspace{-2mm}
\section{Implementation} \label{sec:quicAflNetImpl}
\vspace{-1mm}
The \nameQUICAFLNet{} prototype features two key components: QUIC-AFLNet and Snapfuzz+. QUIC-AFLNet extends \aflnet{} (commit version 6d86ca0c) with the integration of our QUIC-Specific Cryptography Module. 
Snapfuzz+ implements our modifications to Snapfuzz (commit version ef005157), to realise the Synchronisation and Snapshot Protocols. Table~\ref{tab:impl_effort_quic_aflnet} in the Appendix summarises the implementation effort. Here, we describe key aspects of our implementation. 

\vspace{-3mm}
\subsection{QUIC-AFLNet}\label{sec:quic-aflnet_implement}
We follow \aflnet{} guidelines to implement the functions required to parse QUIC input and response sequences. These functions are used to: i)~identify the offset of each input in an input sequence stored in an array and ii) extract state feedback from the \sut{} response. Notably, these functions can handle both encrypted and decrypted message sequences depending on whether the QUIC-specific Cryptographic Module is enabled. In the following, we discuss implementing the QUIC-specific Cryptographic Module.

\vspace{1px}
\noindent\textbf{QUIC-specific Cryptographic Module.~}The module's encryption and decryption process follows the description in~\cite[Section 5]{9001}, with its implementation using helper functions from the OpenSSL-3.0.2~\cite{openssl3} library. It supports testing the \sut{} with TLS1.3 mandatory cipher suite, \texttt{TLS\_AES\_128\_GCM\_SHA256}.

\vspace{1mm}
\noindent\textbf{Seed Corpus.~}Notably, as with \aflnet{}, \nameQUICAFLNet{} saves interesting seeds with a minor modification. When the QUIC-AFLNet identifies an interesting input that leads to a crash, new coverage, or state, it saves \textit{both} the unencrypted input and the respective secrets used for encryption in the results folder. This is important for future mutations of the seed and also enables users to reproduce \sut{} execution using the saved input for further analysis.

\vspace{-4mm}
\subsection{Snapfuzz+}\label{sec:snapfuzz+_impl}

\vspace{-2mm}
We adopt the Snapfuzz framework (see Section~\ref{sec:snapfuzz_background}) and its implementation.  
We extend Snapfuzz's binary re-writing support with our System Call Handle module by implementing the system call handling necessary for QUIC server implementations. Minimising initialisation overhead requires identifying the latest safe point at which to signal the deferred Forkserver to snapshot a target. Our synchronisation protocol extension support the server implementation paradigms employed by QUIC developers. In the following, we briefly detail the modifications made to Snapfuzz to support fuzzing QUIC servers. 

\vspace{2mm}
\noindent\textbf{System Call Handler.~}The QUIC implementations we test rely on an extended set of system calls. In particular, we add support for the following system calls families: \texttt{epoll\_create()}, \texttt{epoll\_ctl()}, \texttt{epoll\_wait()} \texttt{recvmsg()}, \texttt{recvmmsg()}, \texttt{sendmsg()}, and \texttt{sendmmsg()}.

\begin{figure}[!h]
\vspace{-7mm}
\begin{minipage}{0.50\textwidth}
\centering
\begin{minted}
[fontsize=\scriptsize,numbersep=4pt,xleftmargin=10pt,bgcolor=verylightgray,linenos,frame=none]{c}
static int run_server(int fd){
 while(1){
  // ...
  struct msghdr mess = {
   // allocate buffers to store
   // incoming datagram
  };
  ssize_t ret = recvmsg(fd, &mess, 0);
  if(ret > 0){
   // process datagram
  }
  for (i = 0; i != num_conns; ++i){
   // send datagram
  }
 }
}


\end{minted}
\captionof{listing}{Example of a QUIC \textit{receive-send} server main loop.}
\label{lst:other_main_loop}
\end{minipage}
\begin{minipage}{0.50\textwidth}
\centering
\begin{minted}
[fontsize=\scriptsize,numbersep=4pt,xleftmargin=10pt,bgcolor=verylightgray,linenos,frame=none]{c}
static int run_server(int fd){
 while(1){
  // ...
  while(1){
   struct msghdr mess = {
    //allocate buffers to store
    //incoming datagrams
   };
   ssize_t ret = recvmsg(fd, &mess, 0);
   if(ret == -1)
    break;
    //process datagram
  }
  for(i = 0; i != num_conns; ++i){
   //send datagram
  }
 }
}
\end{minted}
\captionof{listing}{Example of a QUIC \textit{receive-break-send} server main loop.}
\label{lst:quicly_main}
\end{minipage}
\vspace{-9mm}
\end{figure}

\vspace{2mm}
\noindent\textbf{Synchronisation Protocol.~}We extended the  synchronisation protocol in Snapfuzz to support the implementation paradigms in QUIC server implementations.

In general, we identified two QUIC server implementation methods. We illustrate these methods in code Listings~\ref{lst:other_main_loop} and \ref{lst:quicly_main}. QUIC implementations employing Listing~\ref{lst:other_main_loop} will always check whether there is data to transmit (line 12--14) after calling a \texttt{recvmsg()} (line 8). The existing synchronisation protocol in Snapfuzz assumes such a design---the server consistently operates in a sequential receive and send order. In contrast, the implementation logic abstracted in code Listing~\ref{lst:quicly_main} repeatedly calls \texttt{recvmsg()} (lines 4--13) until it encounters an error (-1) in the last \texttt{recvmsg()} (line 11). This design choice ensures that the server fully processes all incoming data from the kernel receive buffer. When there is no more data to read, the kernel returns an error (-1) to inform the server that the receive operation will be blocked. At this point, the server enters the sending loop (line 14--16) to transmit its response. However, the existing synchronisation protocol always notifies the fuzzer to deliver input whenever the server invokes a \texttt{recvmsg()}. Consequently, such a server target always has fuzzing input to process and never has a chance to enter the sending loop (lines 14--16).

To support both paradigms, we extend the synchronisation protocol logic to account for the two design choices observed.

\vspace{-2mm}
\section{Evaluation}\label{sec:experiments}
\vspace{-2mm}
In this section, we conduct a series of extensive experiments to evaluate \nameQUICAFLNet{} with other state-of-the-art grey-box network protocol fuzzers. These experiments aim to answer the following research questions:

\vspace{-1mm}
\begin{enumerate}
 \item How does \nameQUICAFLNet{} perform compared to previous state-of-the-art network protocol fuzzers? (Coverage experiments in Section~\ref{sec:experiments_cov})
 \item How does \nameQUICAFLNet{} compare in bug-finding capability to previous state-of-the-art network protocol fuzzers? (Section~\ref{sec:experiments_bugs} and \ref{sec:bug-benchmark1})
 \item How effective are the components and techniques?  (Section~\ref{sec:quic-aflent_stability})
\end{enumerate}

To answer these questions, we first benchmark the performance of the fuzzers based on code coverage metrics. Next, we assess the bug-finding capabilities of the fuzzers by analysing the crashes identified in the first experiment to manually triage all unique bugs. Finally, we assess the impact of the modules we develop for \nameQUICAFLNet{} on fuzzing performance using code coverage metrics. 

\vspace{-1mm}
\subsection{Experimental Setup}\label{sec:experiments_setup} 
\vspace{-1mm}
All of the experiments are run on an AMD Ryzen Threadripper 3990X CPU with 256 GB of RAM where one core is allocated per fuzzing instance. 

We compare \nameQUICAFLNet{} with three\footnote{We also considered Bleem~\cite {luo2023bleem}, which is not publicly available, and contacted the authors to request source code  for benchmarking, we have not received a reply.} 
state-of-the-art, open-source, grey-box fuzzers: \fn{}~\cite{fuzztrution-net}, \chatafl{}~\cite{chatafl}, and \aflnet{}~\cite{aflnet} (baseline). These are selected for the following reasons: i) \fn{} outperformed other grey-box fuzzers (\textsc{SGFuzz}~\cite{SGF} and \textsc{StateALF}\cite{natella2022stateafl}) in fuzzing secure protocols; ii) \chatafl{} is a concurrent work not included in the \fn{} benchmark and represents the first fuzzer to employ an LLM in network protocol fuzzing; iii) \aflnet{} serves as a baseline since ours is built on top of \aflnet{} whilst also serving to support our ablations study; and iv) they are all open-sourced.  All fuzzers are run with Address Sanitiser enabled to detect bugs that do not result in a crash (all configuration settings are  
in Appendix~\ref{apd:fuzzer-settings}).

\begin{figure*}[b!]
  \centering
  \includegraphics[width=1.0\linewidth, trim={120 0 120 50, clip}]{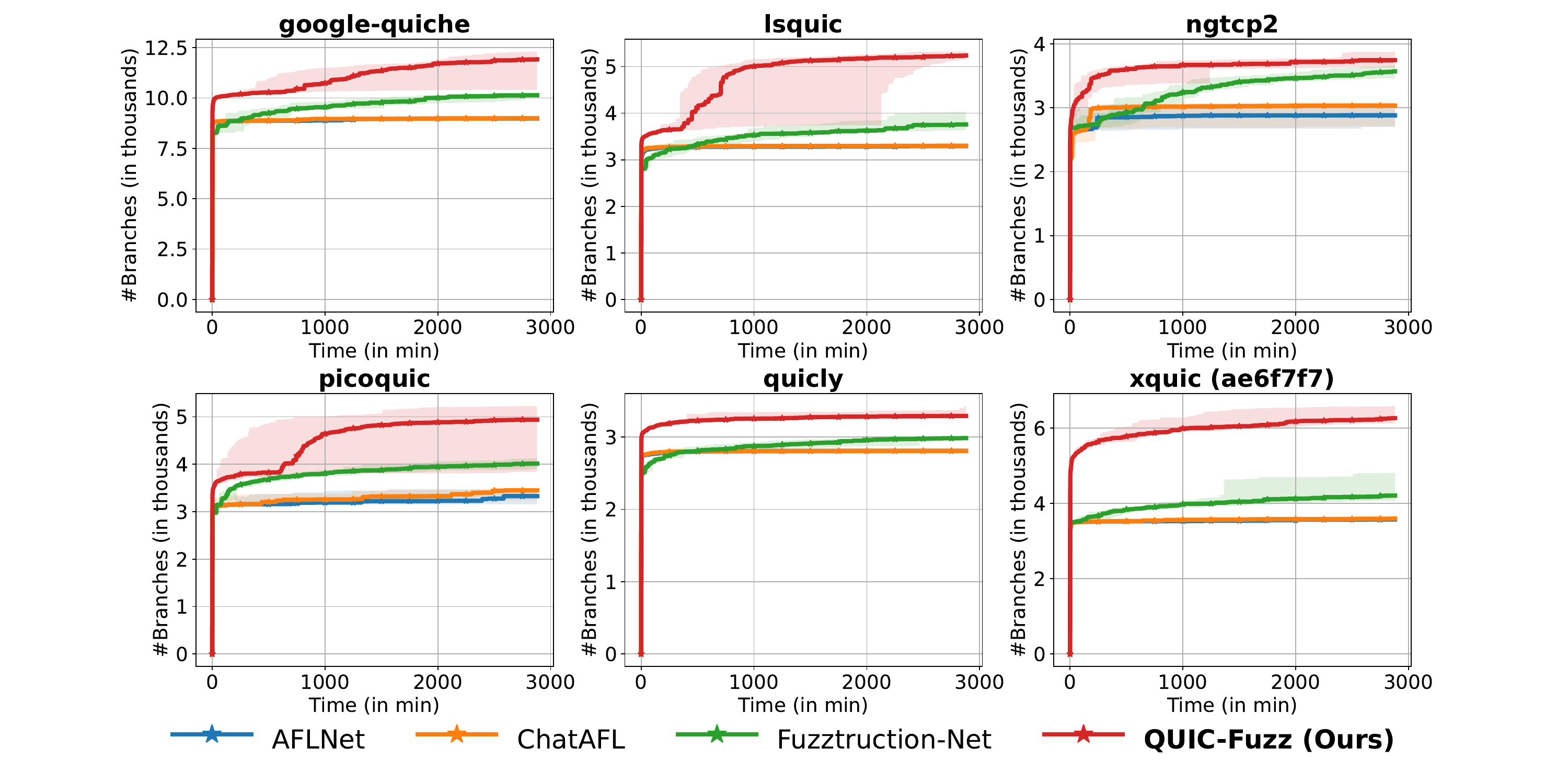}
  \vspace{-8mm}
  \caption{The median branch coverage (in thousands) achieved by fuzzers in ten trials of 48-hour executions. The coloured area represents the best/worst coverage achieved by a fuzzer. Notably, we fuzzed two versions of \tgt{Xquic} (Alibaba), with the coverage reported here corresponding to the version we fuzzed before identifying the bugs we reported. The subsequent version (released with the fixes for the bug we identified) was found to contain a new bug, as Table~\ref{tab:quic_aflnet_faults} shows.} 
  \label{fig:quic-aflnet-cov}
  \vspace{-2mm}
\end{figure*} 

\vspace{2mm}
\noindent\textbf{QUIC Server Targets}~(\sut{}s).~We select \textit{publicly} available and \textit{well-maintained} QUIC servers for our evaluations. We summarise the servers we selected in Table~\ref{tab:quic-aflnet_server_tested}. We test \quicAflnetImpCount{} QUIC implementations written in C/C++. These include \tgt{Google-quiche} from Google, \tgt{XQUIC} from Alibaba, and \tgt{LSQUIC} from Lite Speed Technologies. For replay-based fuzzers (\nameQUICAFLNet{}, \chatafl{}, \aflnet{}), we patch static values for secrets, verification data, connection IDs, and packet numbers in the \sut{}s to eliminate non-determinism caused by ephemeral values.

\vspace{-4mm}
\subsection{Coverage Experiments}\label{sec:experiments_cov}
\vspace{-2mm}
\begin{mdframed}[backgroundcolor=black!5,rightline=false,leftline=false,topline=false,bottomline=false,roundcorner=1mm,everyline=false]
Coverage results are reported in Figure~\ref{fig:quic-aflnet-cov}. In summary, \nameQUICAFLNet{} outperforms all others  and we can observe an increase in branch coverage of up to 84\%. Further, Table~\ref{tab:other_fuzzzer_stat} (in Appendix) shows that \nameQUICAFLNet{} achieves \textit{median} code coverage improvements that are statistically significant across all targets compared to the \aflnet{} and \chatafl. We observe significant improvements in 5 out of 6 targets compared to \fn{}.
\end{mdframed}

To evaluate the effectiveness of a fuzzer, we use code coverage metrics to provide a quantitative comparison. We conduct a fuzzing campaign that involves running each fuzzer for 48 hours for each of the \quicAflnetImpCount{} targets shown in Table~\ref{tab:quic-aflnet_server_tested}. To remove randomness, we conduct ten trials~\cite{klees2018evaluating} of each experiment.

Generally, we observe that both \aflnet{} and \chatafl{} saturate coverage early and fail to explore new code to achieve higher coverage. As discussed in \Circled{C1} of Section~\ref{sec:quicAflNetDesign}, because \chatafl{} and \aflnet{} apply mutations directly to the seed, most of the mutated inputs cannot pass integrity checks and are discarded without further processing. Consequently, these fuzzers are limited to exploring the functions that deal with unencrypted components of a packet and those with the correct integrity (the initial seed to the fuzzer). Further,  \chatafl{} and \aflnet{} do not synchronise fuzzing events, hence, performance is   impacted by the non-determinism issues discussed in \Circled{C2}. As a result, the fuzzers cannot reliably re-use an interesting test cases to make incremental progress.

In contrast, \fn{} uses a QUIC client harness that can account for the negotiated encryption secrets when generating fuzz inputs. This harness effectively provides a built-in secure and liable transport layer with the \sut{}, mitigating the need to inject delays that characterises the poor performance of \aflnet{} and ChatALF while avoiding problems with non-determinism.  

However, \nameQUICAFLNet{} performs better than \fn{} in terms of coverage. During the experiment, the throughput of \fn{} was observed to be significantly lower ($\sim0.20$ executions per second) than that of \nameQUICAFLNet{} ($\sim25$ executions per second). We attribute the performance difference to: i) the time needed to perform dynamic fault injection on the client harness, ii) the initialisation overhead of the \sut{}---which \nameQUICAFLNet{} mitigates by employing a Snapshot Protocol---as well as the additional overhead of initialising the client harness (see \Circled{C3}), and iii) because the harness is effectively a complete client application and therefore expends time processing packets and their payloads, unlike \nameQUICAFLNet{}, which is purpose-built for fuzzing.

\begin{table*}[]
\vspace{-8mm}
\centering
\caption{Overview of the software bugs discovered in the fuzzing campaign with \nameQUICAFLNet{}. M11 is a known bug that is not addressed at the time of testing.}
\label{tab:quic_aflnet_faults}
\resizebox{\textwidth}{!}{%
\begin{tabular}{@{}p{0.2\textwidth}p{0.80\textwidth}ccp{0.15\textwidth}@{}}
\toprule
Server   &  Fault Description   & Bug ID              & Disclosed         & Status\\ \toprule
\tgt{Picoquic} & \textsf{CVE-2024-45402}: Double Free in Picotls when processing an unexpected public key length. & M1 & \checkmark & Fixed.  \\
 & NULL pointer dereference when copying data from a STREAM frame.
& M2 &   \checkmark   & Fixed. \\ 
 & Heap OOB read on a zero-sized heap when parsing a QPACK header.
& M3 &   \checkmark   & Fixed. \\
 & Null pointer dereference after failing to match a stream ID with an existing stream context.
& M4 &   \checkmark   & Fixed. \\ \midrule
\tgt{Quicly}   & \textsf{CVE-2024-45396}: Assertion failure when processing an ACK frame in the draining stage.                                                        & M5                  &        \checkmark           & Fixed.                              \\ \midrule
\tgt{Xquic}~(\texttt{ae6f7f7})   & NULL pointer dereference when handling a Stateless Reset packet on a destroyed connection.                                                 & M6                  &       \checkmark            & Fixed.                              \\ 
        &  Stack OOB read when copying a string for logs.                                                                                                   & M7                  &       \checkmark            &       Reported.        
        \\ 
        &  NULL pointer dereference when checking whether application data is available for reading or when handling the FIN bit in the byte stream.                                                                                                   & M8                  &       \checkmark            &       Reported.             \\
         & Heap OOB read when comparing an HTTP/3 header.                                                                                                   & M9                  &       \checkmark            &       Reported.             \\ \midrule
 \tgt{Xquic}~(\texttt{6803065})   & NULL pointer dereference when logging a Version Negotiation event.                                                                                                   & M10                  &       \checkmark            & Fixed.                   \\ \midrule
Ngtcp2   & Assertion failure when writing a CONNECTION\_CLOSE frame using an unsupported cipher suite by the TLS library (known bug).                                                                                                    & M11                  &       \checkmark            & Fixed.                   \\ 
 
 \bottomrule          
\end{tabular}%
}
\vspace{-2mm}
\end{table*}

\vspace{-8mm}
\subsection{Bug-Finding Capability and Case Studies}\label{sec:experiments_bugs} 

\begin{mdframed}[backgroundcolor=black!5,rightline=false,leftline=false,topline=false,bottomline=false,roundcorner=1mm,everyline=false]
\nameQUICAFLNet{} discovered \bugcountQUICAFLNet{} new security vulnerabilities (as summarised in Table~\ref{tab:quic_aflnet_faults}). These discoveries have resulted in a bug bounty award and 2 CVE assignments thus far. We responsibly disclosed all vulnerabilities to the respective developers. We detail Proofs of Concept (PoCs) for reproducing  bugs and bug \textit{\textbf{case studies}} with bug \textit{impact} discussions at our GitHub repo \textsf{\href{https://github.com/QUICTester/QUIC-Fuzz}{https://github.com/ QUICTester/QUIC-Fuzz}} and Appendix~\ref{apd:case-studies}. 
\end{mdframed}

\subsection{Bug Benchmark}\label{sec:bug-benchmark1} 
\begin{figure*}[t!]
  \centering
  \includegraphics[width=1.0\linewidth, trim={0 20 0 20, clip}]{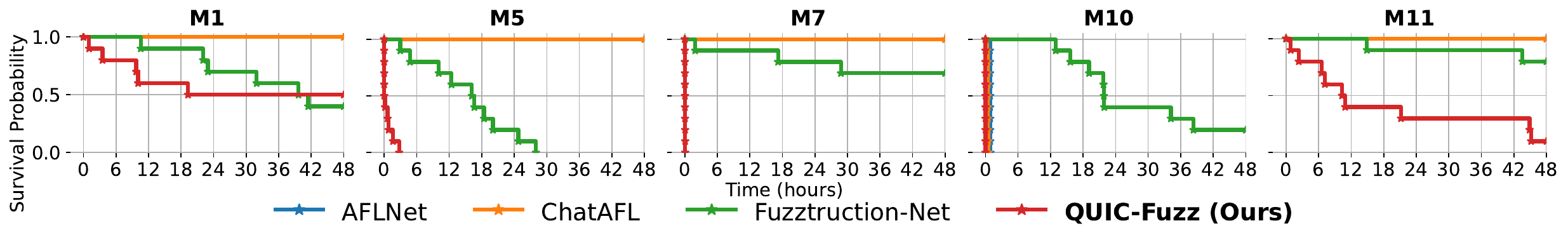}
  \caption{Survival probability plots for the four bugs (M1, M5, M7, M10, M11 described in Table~\ref{tab:quic_aflnet_faults}) found by at least two of the benchmark fuzzers within 48 hours, as shown in Table~\ref{tab:quic_aflnet_bug_bench}. Only M10 is found by all of the fuzzers.
  } 
  \label{fig:quic-aflnet-survival-plot}
  \vspace{-6mm}
\end{figure*}

\begin{mdframed}[backgroundcolor=black!5,rightline=false,leftline=false,topline=false,bottomline=false,roundcorner=1mm,everyline=false]
As shown in Table~\ref{tab:quic_aflnet_bug_bench}, \bugcountQAlfonly{} out of \bugcountQUICAFLNet{} bugs can only be discovered by \nameQUICAFLNet{}. In addition, \nameQUICAFLNet{} triggers those bugs that can be found by at least two fuzzers more reliably and with significantly less time  (see Figure~\ref{fig:quic-aflnet-survival-plot}). 
\end{mdframed}

\begin{wraptable}{l}{0.5\textwidth} 
\centering
\vspace{-20pt}
\caption{Unique bug benchmark study results from a fuzzing campaign of ten trials of 48-hour fuzzing runs.}
\label{tab:quic_aflnet_bug_bench}
\resizebox{0.5\textwidth}{!}{%
\begin{tabular}{@{}lcccc@{}}
\toprule
Bug ID & AFLNet & ChatAFL & Fuzztruction-Net & \nameQUICAFLNet{} \\ \toprule
M1    &   \textcolor[rgb]{0.7,0.7,0.7}{0/10}     &   \textcolor[rgb]{0.7,0.7,0.7}{0/10}      &       \textbf{6/10}           &     5/10        \\ \midrule
M2    &   \textcolor[rgb]{0.7,0.7,0.7}{0/10}     &   \textcolor[rgb]{0.7,0.7,0.7}{0/10}      &       \textcolor[rgb]{0.7,0.7,0.7}{0/10}           &     \textbf{3/10}        \\ \midrule
M3    &   \textcolor[rgb]{0.7,0.7,0.7}{0/10}     &   \textcolor[rgb]{0.7,0.7,0.7}{0/10}      &       \textcolor[rgb]{0.7,0.7,0.7}{0/10}           &     \textbf{2/10}        \\ \midrule
M4    &   \textcolor[rgb]{0.7,0.7,0.7}{0/10}     &   \textcolor[rgb]{0.7,0.7,0.7}{0/10}      &       \textcolor[rgb]{0.7,0.7,0.7}{0/10}           &     \textbf{1/10}        \\ \midrule
M5    &   \textcolor[rgb]{0.7,0.7,0.7}{0/10}     &   \textcolor[rgb]{0.7,0.7,0.7}{0/10}      &       \textbf{10/10}           &     \textbf{10/10}        \\ \midrule
M6    &   \textcolor[rgb]{0.7,0.7,0.7}{0/10}     &   \textcolor[rgb]{0.7,0.7,0.7}{0/10}      &       \textcolor[rgb]{0.7,0.7,0.7}{0/10}           &     \textbf{1/10}        \\ \midrule
M7    &   \textcolor[rgb]{0.7,0.7,0.7}{0/10}     &   \textcolor[rgb]{0.7,0.7,0.7}{0/10}      &       3/10           &     \textbf{10/10}        \\ \midrule
M8    &   \textcolor[rgb]{0.7,0.7,0.7}{0/10}     &   \textcolor[rgb]{0.7,0.7,0.7}{0/10}      &       \textcolor[rgb]{0.7,0.7,0.7}{0/10}           &     \textbf{10/10}        \\ \midrule
M9    &   \textcolor[rgb]{0.7,0.7,0.7}{0/10}     &   \textcolor[rgb]{0.7,0.7,0.7}{0/10}      &       \textcolor[rgb]{0.7,0.7,0.7}{0/10}           &     \textbf{2/10}        \\ \midrule
M10    &  \textbf{10/10}     &   \textbf{10/10}      &       3/10           &     \textbf{10/10}        \\ \midrule
M11    &  \textcolor[rgb]{0.7,0.7,0.7}{0/10}     &   \textcolor[rgb]{0.7,0.7,0.7}{0/10}      &       2/10           &     \textbf{9/10}        \\ 
\bottomrule
\end{tabular}%
}
\vspace{-13pt}
\end{wraptable}

To measure the bug-finding capability of \nameQUICAFLNet{} with other fuzzers, we consider a bug-based benchmark and conduct two experiments. We manually triaged the crashes identified by all of the fuzzers during the coverage experiments in Section~\ref{sec:experiments_cov} and spent efforts to identify the unique bugs (new and known bugs) uncovered by each fuzzer. First, we use data to compare the probability of a unique bug being triggered within ten fuzzing trials. We summarise the results in Table~\ref{tab:quic_aflnet_bug_bench}. Second, to assess the efficiency of each fuzzer and the reliability of its bug-discovering capability, we adopt \cite{wagner2017elastic}'s approach, using the Kaplan-Meier estimator~\cite{kaplan1958nonparametric} to model the \textit{survival function} of those bugs that are discovered by at least \textit{two} of the fuzzers. We report the results in Figure~\ref{fig:quic-aflnet-survival-plot}. We discuss the results in detail in the following sections.

\vspace{2mm}
\noindent\textbf{Unique Bug Benchmark.~} As shown in Table~\ref{tab:quic_aflnet_bug_bench}, the second-best fuzzer, \fn{}, detected only 5 out of 11 of the bugs uncovered by \nameQUICAFLNet. Overall, for three of the five common bugs uncovered by both \nameQUICAFLNet{} and \fn{}, we can observe the probability of the bugs being discovered by \nameQUICAFLNet{} to be significantly higher than \fn.

As expected, both ALFNet and \chatafl{} perform poorly, indicating the need to address the challenges we described as well as the need for a protocol-specific fuzzer that is effective at discovering bugs. Interestingly, \chatafl{} and \aflnet{} are able to discover one bug, \textbf{M10}. This particular bug can be triggered by mutating the version field in the Initial packet header. Notably, the version field starts at the second byte of the Initial packet and is unprotected. When a server observes a version it does not support, it immediately sends a Version Negotiation packet without further processing the remaining packet. Consequently, a mutation to a seed---as performed in \aflnet{}---can trigger \textbf{M10}, as we explain in Section~\ref{sec:experiments_bugs}.  

\vspace{2mm}
\noindent\textbf{Survival Probability.~}Overall, \nameQUICAFLNet{} can detect bugs more quickly and consistently than other fuzzers: it identifies three out of the five common bugs (\textbf{M5}, \textbf{M7}, and \textbf{M10}) with a 100\% success rate within 4 hours of fuzzing. In contrast, the second-best fuzzer, \fn{}, is guaranteed (100\%) to find only one bug (\textbf{M5}) within 29 hours of fuzzing. 

Interestingly, \nameQUICAFLNet{}, \chatafl{}, and \aflnet{} can detect \textbf{M10} within 5 minutes in all ten trials (100\% success). By comparison, \fn{} demonstrates only a 80\% chance of triggering the bug within 48 hours of fuzzing. We attribute this to i)~the overheads impacting fuzzing throughput as we discussed in Section~\ref{sec:experiments_cov} and ii)~the \fn{} scheduling algorithm, as explained  in~\cite{fuzztrution-net}, also not favouring inputs that covers less code. This may prevent \fn{} from exercising code paths that do \textit{not} involve cryptographic operations or deep functionality, where bugs like \textbf{M10} can reside.

\begin{table*}[]
\vspace{-8mm}
\centering
\caption{Median branch coverage after ten 48-hour runs with different modules enabled in \nameQUICAFLNet{}. The modules are added from the left (AFLNet as the baseline) to the right (+Snapshot Protocol: the \textit{full} version of \nameQUICAFLNet{}). The percentages show the change in coverage compared to the previous configuration. Changes \textless{}0.1\% are not displayed and statistically significant changes are marked in bold (based on a Mann-Whitney U test at a 0.05 significance level).}
\label{tab:quic-aflnet_ablation_study}
\resizebox{\textwidth}{!}{%
\begin{tabular}{l|c|rlc|rlc|rlc}
\cline{1-11}
\multirow{2}{*}{QUIC server} \rule{0pt}{10pt}  & \multicolumn{1}{c|}{AFLNet (baseline)} & \multicolumn{3}{c|}{+QUIC-specific Cryptographic Module} & \multicolumn{3}{c|}{+Synchronisation   Protocol}              & \multicolumn{3}{c}{+Snapshot   Protocol (\textbf{\nameQUICAFLNet{}})} \\
                             &                   &      &              & \textit{p-value}   &  &         & \textit{p-value} &  &         & \textit{p-value} \\ \cline{1-11}
\tgt{Google-quiche}  \rule{0pt}{10pt}   & 8984            & 10354           & \textbf{(+15\%)}      & 0.0002       & 11545         & \textbf{(+12\%)} & 0.0058             & 11915                       & (+3\%)           & \textcolor[rgb]{0.7,0.7,0.7}{0.1405}  \\
\tgt{Lsquic}                       & 3294                 & 3584            & \textbf{(+9\%)}       & 0.0002       & 4724          & \textbf{(+32\%)} & 0.0002             & 5238                        & \textbf{(+11\%)} & 0.0003  \\
\tgt{Ngtcp2}                       & 2881                 & 3779            & \textbf{(+31\%)}      & 0.0002       & 3711          & (-2\%)           & \textcolor[rgb]{0.7,0.7,0.7}{0.3447}             & 3744      & (+1\%)           & \textcolor[rgb]{0.7,0.7,0.7}{0.3447}  \\
\tgt{Picoquic}                     & 3328                 & 3805            & \textbf{(+14\%)}      & 0.0002       & 4891          & \textbf{(+29\%)} & 0.0009             & 4940                        & (+1\%)           & \textcolor[rgb]{0.7,0.7,0.7}{0.6232}  \\
\tgt{Quicly}                       & 2807                 & 3260            & \textbf{(+16\%)}      & 0.0002       & 3291          & \textbf{(+1\%)}  & 0.0046             & 3291                        &                  & \textcolor[rgb]{0.7,0.7,0.7}{0.7334}  \\
\tgt{Xquic}                        & 3568                 & 5784            & \textbf{(+62\%)}      & 0.0002       & 6211          & \textbf{(+7\%)}  & 0.0002             & 6262                        & (+1\%)           & \textcolor[rgb]{0.7,0.7,0.7}{0.3075} \\ \cline{1-11}
\end{tabular}%
}
\end{table*}

\vspace{-10mm}
\subsection{Effectiveness of \nameQUICAFLNet{} Modules}\label{sec:quic-aflent_stability} 

\begin{mdframed}[backgroundcolor=black!5,rightline=false,leftline=false,topline=false,bottomline=false,roundcorner=1mm,everyline=false]
We report the ablation study results in Table~\ref{tab:quic-aflnet_ablation_study}. We can observe each technique applied in \nameQUICAFLNet{} is effective and contributes to improving performance in a manner that is statistically significant. 
\end{mdframed}

\vspace{-2mm}
To evaluate the effectiveness of the techniques within the three modules that we investigate using \nameQUICAFLNet{}, we conduct three additional series of experiments, one for each module. Specifically, we run \nameQUICAFLNet{} on all of the targets with different subsets of modules enabled and collect the resulting code coverage. We use \aflnet{} as our baseline. We begin with the configuration where only the QUIC-specific Cryptographic Module is activated. Then, we sequentially enable the Synchronisation Protocol and Snapshot Protocol. Notably, we evaluate whether each addition has a statistically significant impact on the branch coverage achieved by the fuzzer in ten trials. We report our results in Table~\ref{tab:quic-aflnet_ablation_study} and defer the discussing the effectiveness of each module to Appendix~\ref{apd:ablation}.

\vspace{-2mm}
\section{Discussion} \label{sec:discussion}
\vspace{-2mm}
In this section, we discuss the potential limitations of our design and provide insights for future research on fuzzing secure protocols.

\vspace{1mm}
\noindent\textbf{Ephemeral Values.~} 
Manual configuration is needed in \nameQUICAFLNet{} to address non-determinism that can emerge from ephemeral values, such as Connection IDs used in connections in QUIC over a UDP data stream. As explained in Section~\ref{sec:experiments_setup}, these fields are patched with static values to ensure the \sut{} always uses the same values in each fuzzing iteration. Notably, \fn{} can remove this effort because the fault injection method employs a complete client application able to establish a secure connection with a target.

\vspace{1mm}
\noindent\textbf{System Call Handler.~}
\nameQUICAFLNet{} does not currently support Rust servers (Neqo, Quiche, Quinn, S2n-quic). Fuzzing these additional servers---while meeting our selection criteria of being well maintained and publicly available---requires extending binary re-writing support for system calls within the System Call Handler to those used in the target list of servers. We leave this for future studies.

\vspace{1mm}
\noindent\textbf{Grammar Aware Mutations.~}
\nameQUICAFLNet{} performs random mutations on decrypted packet content without considering packet structure and grammar. This suggests that one avenue for improving \nameQUICAFLNet{} would involve enabling grammar-based mutations. Notably, \chatafl{} demonstrates the effectiveness of such an approach for \textit{non-secure} application layer protocols. However, in our adaptation of \chatafl{} to fuzz QUIC, we observed that LLMs---specifically GPT3.5, which is used by \chatafl---lack the ability to extract grammar when interpreting complex transport layer protocol messages, undermining their effectiveness for the task. This provides avenues for further exploration of the use of LLMs for fuzzing transport layer protocols.

\vspace{-2mm}
\section{Related Work} \label{sec:quic_aflnet_relatedWork}
\vspace{-2mm}
Our work investigates fuzzing for  security testing QUIC. Here, we discuss related approaches for fuzzing QUIC and network protocol fuzzing, more generally.

\vspace{1mm}
\noindent\textbf{QUIC Testing.~}Researchers have explored testing methods to check the adherence of a QUIC implementation to its specification, for versions \textit{before} its ratification by the IETF, have also been developed~\cite{ferreira2021prognosis,crochet2021verifying,mcmillan2019formal,rath2018interoperability,goel2020testing,quictester}.

\vspace{1mm}
\noindent\textbf{Protocol Fuzzing.~}In general, grey-box fuzzers for network protocols and applications~\cite{aflnet,SGF,natella2022stateafl,nyxnet,snapFuzz,qin2023nsfuzz,li2022snpsfuzzer} operate in a feedback-driven loop to generate fuzz input by mutating packets captured from previous network sessions. Recent \chatafl{}~\cite{chatafl}, leverages pre-trained LLMs to perform grammar-aware mutations and predict the next input when testing \textit{application layer} protocols to generate more effective inputs. Other than grey-box fuzzing, there are approaches to fuzz test network protocols, such as black-box fuzzing~\cite{peach-fuzzer,boofuzz,gascon2015pulsar,luo2023bleem,feng2021snipuzz,wu2024logos}, symbolic executions~\cite{chen2020savior,ognawala2018improving,ff2bb9e7c53b40609a4218d1a7df2100} and grammar-based fuzzing~\cite{aschermann2019nautilus,wang2019superion,chatafl}.
By contrast, \nameQUICAFLNet{} focuses on generating mutated yet integrity-protected packets. 

\vspace{1mm}
\noindent\textbf{Fuzzing QUIC.~}While we are \textit{unaware} of an effective grey-box fuzzer for the \textit{IETF-ratified QUIC}, fuzzing techniques and tools~\cite{aflnet,natella2022stateafl,luo2023bleem,fuzztrution-net} capable of fuzz-testing QUIC exist. \textsc{StateAFL}, an extension of \aflnet{}~\cite{aflnet}, infers the SUT states by observing the content of SUT’s memory allocations to eliminate the need to write a custom message parser. 
\textsc{Bleem}~\cite{luo2023bleem} introduces a Man-in-the-Middle approach that fuzzes both the client and the server simultaneously by actively intercepting and modifying packets exchanged. However, \aflnet{}, \textsc{StateAFL}, and \textsc{Bleem} does not account for packet protections (such as encryption) in secure protocols like QUIC; rather, they perform direct mutations on encrypted packets (seeds) and suffer from the problems discussed in Section~\ref{sec:quicCryptoModule}.

A concurrent study, \fn{}~\cite{fuzztrution-net}, leverages a network peer (client or server) as a harness to generate fuzzing inputs.  It dynamically injects faults into the harness to: i) modify variable values for targeted field mutations, and ii) alter the harness's execution trace (e.g., changing the function call destination) to achieve sequence mutation. As discussed in Section~\ref{sec:discussion}, \fn{} does not require protocol-specific crypto modules and deals with ephemeral values without manual patching. But, the approach requires identifying the optimal fault insertion location to ensure the peer remains operational while testing. Further, the number of variables available for mutation in a harness can exceed the number of fields in a QUIC packet. Hence, \fn{} has a larger exploration space during mutation compared to a replay-based fuzzer like \nameQUICAFLNet{} and must carefully select impactful variables for mutation. Consequently, \nameQUICAFLNet{}'s execution speed is 125\% faster than \fn{} (25 exec/s vs 0.2 exec/s on our execution hardware). 

\vspace{1mm}
\noindent\textbf{Optimising the Fuzzing Loop.~}Previous works~\cite{zeng2020multifuzz,prenny} have also intercept network system calls to synchronise fuzzing events. In addition, snapshot techniques~\cite{xu2017designing,dong2020time,li2022snpsfuzzer,nyxnet} have been applied to fuzzing to mitigate initialisation overhead, especially on stateful targets that require reset~\cite{aflnet}. Because Snapfuzz~\cite{snapFuzz} adopts both of these methods---synchronisation and snapshotting---for \aflnet{}, we adopted and extended it in our work.

\vspace{-3mm}
\section{Conclusions and Future Work} \label{sec:conclusion}
\vspace{-3mm}
In this study, we have developed and implemented a grey-box mutation-based fuzzer---\nameQUICAFLNet---to discover vulnerabilities beyond specification bugs in QUIC server-side implementations. We have evaluated \nameQUICAFLNet{} on six QUIC servers and compared the fuzzer developed against three state-of-the-art network protocol fuzzers. Extensive evaluations demonstrate that \nameQUICAFLNet{} is both efficient and effective for fuzzing QUIC targets. In particular, \nameQUICAFLNet{} outperforms all other fuzzers---with up to an 84\% increase in code coverage---and uncovers \bugcountQUICAFLNet{} security vulnerabilities, with two CVEs assigned from developers and a bug bounty reward received. Extending \nameQUICAFLNet{} to leverage LLMs to perform grammar-aware mutations and predict the next message during fuzzing represent potentially fruitful avenues for future research.

\bibliographystyle{splncs04} 
\bibliography{references}   

\appendix
\section{Appendix}

\subsection{ALFNet Background}\label{apd:aflnet}

\begin{figure}[t!]
  \centering
  \includegraphics[width=0.7\linewidth]{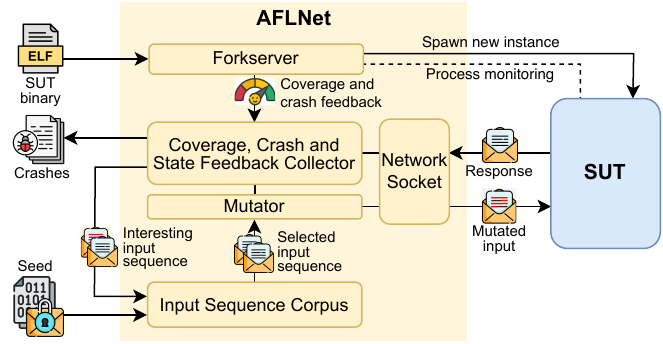}
  \caption{An overview of \aflnet{}.} 
  \label{fig:AFLNet}
\end{figure}

\aflnet{}~\cite{aflnet}, shown in Figure~\ref{fig:AFLNet}, is a grey-box mutation-based fuzzer specifically designed for stateful network protocols. From a systems perspective, \aflnet{} operates as the client application and directs its generated inputs in a sequence through network sockets (e.g. TCP/IP, UDP/IP) to the SUT. The inputs are generated by mutating a seed (sequences of recorded message exchanges between a client and server). Differing from traditional fuzzers (e.g. AFL), \aflnet{} introduces the concept of using the SUT's state feedback to guide the fuzzing process. It uses a modified form of edge coverage combined with state awareness to efficiently identify interesting inputs that alter the control flow of the target application to help explore new code and states. 
In the following, we briefly describe the events that occur in the fuzzing loop.

In each fuzzing iteration, \aflnet{} selects an input sequence from the Input Sequence Corpus to mutate. After the mutation process, \aflnet{} requests a new SUT instance from the Forkserver and waits for a pre-determined delay to ensure that the SUT is initialised and ready to consume inputs. Then, \aflnet{} sends input sequences to the SUT. To ensure synchronisation of communications, \aflnet{} adds user-specified communication delays between inputs. If \aflnet{} receives a response from the SUT within these delays, it processes the response to extract state information---for example, response codes or message types---to dynamically infer the SUT's state machine. During the execution of the target, \aflnet{} also collects code coverage. Once the input sequence is consumed by the SUT, the fuzzer sends a \texttt{SIGTERM} signal to terminate the SUT. If the input sequence uncovers new execution paths (edge coverage) or reaches unexplored protocol states, it is added to the Input Sequence Corpus for further exploration.

Notably, \aflnet{} includes a Forkserver mode to reduce the loading and initialisation time for an SUT. Before \aflnet{} enters the fuzzing loop, it creates a Forkserver that loads the SUT with the \texttt{execve()} system call and snapshots the SUT immediately before it executes the main function. This allows \aflnet{} to quickly use the Forkserver to \texttt{fork()} a new SUT instance from the snapshot in each fuzzing iteration. In addition, \aflnet{} offers an optional deferred Forkserver mode to further reduce the SUT initialisation overhead. In this mode, users must manually add the \texttt{\_\_AFL\_INIT()} function call in the SUT’s source code to allow the fuzzer to defer the creation of the Forkserver until the specified point in the source code.

Importantly, developing a fuzzing harness (the code needed to fuzz a target) is not straightforward in the context of \aflnet{}, let alone for a new protocol. Developing a harness includes the client code, determining delay configurations, and the clean-up scripts. In our work, we consider the optimisation proposed by SnapFuzz to simplify the creation of fuzzing harnesses and adopt it to address the challenges associated with building an effective fuzzer for QUIC.

\subsection{Configurations Used in Benchmarks and Implementation Effort}\label{apd:fuzzer-settings}

\noindent\textbf{\aflnet}~(commit \texttt{6d86ca0c}).~We follow their description to enable state-aware mode
and the region-level mutation operators. As in \nameQUICAFLNet{}, we include the required functions for parsing QUIC input and response sequences.

\noindent\textbf{\chatafl}~(commit \texttt{1ea603eb}).~Because \chatafl{} is built on top of \aflnet{}, we include the same functions we added to \aflnet{} and enable the same fuzzing configurations. To ensure that the LLM understands the QUIC protocol contexts, we modify the LLM prompt templates in \chatafl{} by replacing all the "request" keywords with "packet" because QUIC does not operate under the concept of requests. 

\noindent\textbf{\fn}~(commit \texttt{9e981022}).~\fn{} generates fuzz inputs by injecting faults into clients. As such, we use the \tgt{Ngtcp2} QUIC client (commit version \texttt{f3f15b6}) as the harness because it was the chosen client in the benchmark with QUIC servers tested in \fn{}. 

\vspace{2mm}
We detail the QUIC server implementations tested in our experiments in Table~\ref{tab:quic-aflnet_server_tested} and summarise the implementation effort in Table~\ref{tab:impl_effort_quic_aflnet}.

\begin{table}[ht]
\vspace{-6mm}
\centering
\begin{minipage}{0.66\textwidth}
\caption{QUIC server implementations tested.}
\label{tab:quic-aflnet_server_tested}
\resizebox{\linewidth}{!}{%
\begin{tabular}{lccl}
\toprule
\textbf{Name}          & \textbf{Commit Version} & \textbf{Language} & \multicolumn{1}{c}{\textbf{URL}}                     \\ \toprule
\tgt{Google-Quiche} & \texttt{149b7e6}        & C++      & \url{https://github.com/google/quiche}            \\
\tgt{Lsquic}        & \texttt{c4f359f}        & C        & \url{https://github.com/litespeedtech/lsquic}     \\
\tgt{Ngtcp2}        &\texttt{e2372a8}       & C/C++    & \url{https://github.com/ngtcp2/ngtcp2}            \\
\tgt{Picoquic}      & \texttt{8f4f77f}        & C        & \url{https://github.com/private-octopus/picoquic} \\
\tgt{Quicly}        & \texttt{6a90372}        & C        & \url{https://github.com/h2o/quicly}               \\
\tgt{Xquic}         & \texttt{ae6f7f7}, \texttt{6803065}        & C        & \url{https://github.com/alibaba/xquic}           \\ \toprule
\end{tabular}%
}
\end{minipage}%
\hfill
\vline%
\hfill
\begin{minipage}{0.32\textwidth}
\centering
\caption{Implementation effort.}
\label{tab:impl_effort_quic_aflnet}
\resizebox{\linewidth}{!}{%
\begin{tabular}{lcr}
\toprule
\textbf{Component}                         & \textbf{Library}       & \multirow{2}{*}{\textbf{\shortstack{Lines of \\ Code}}} \\
& & \\ \toprule
\multirow{2}{*}{QUIC-AFLNet} & AFLNet, & \multirow{2}{*}{3530}          \\
 & OpenSSL-3.0.2 &           \\ \midrule
Snapfuzz+                         & Snapfuzz      & 724          \\
\bottomrule
\end{tabular}%
}
\end{minipage}
\vspace{-9mm}
\end{table}

\subsection{Statistical Significance Testing of Branch Coverage}\label{apd:stat-testing}

Coverage results are reported in Figure~\ref{fig:quic-aflnet-cov}. In summary, \nameQUICAFLNet{} outperforms all others fuzzers and we can observe an increase in code coverage of up to 84\%. Table~\ref{tab:other_fuzzzer_stat} reports that \nameQUICAFLNet{} achieves \textit{median} code coverage improvements that are statistically significant across all targets compared to the \aflnet{} and \chatafl. We observe significant improvements in 5 out of 6 targets compared to \fn{}.
\begin{table*}[h!]
\vspace{-6mm}
\centering
\caption{The median branch coverage achieved by our \nameQUICAFLNet{} compared to \aflnet{}, \chatafl{} and \fn{} in ten trials of 48-hour executions. The percentage shows the improvement of \nameQUICAFLNet{} in branch coverage compared to the SToA fuzzer. The statistically significant changes are marked in bold (based on a Mann-Whitney U test with a 0.05 significance threshold).}
\label{tab:other_fuzzzer_stat}
\resizebox{\textwidth}{!}{%
\begin{tabular}{l|c|rlr|rlr|rlr}
 \cline{1-11}
\multirow{2}{*}{QUIC server} \rule{0pt}{10pt} & \textbf{\nameQUICAFLNet{} (Ours)} & \multicolumn{3}{c|}{\aflnet{}} & \multicolumn{3}{c|}{\chatafl{}} & \multicolumn{3}{c}{\fn{}} \\
                             &                  &               &                       & \textit{p-value}         &           &                    & \textit{p-value}      &               &                       & \textit{p-value}        \\ \cline{1-11}
\tgt{Google-Quiche}    \rule{0pt}{10pt} & 11915            & 8984          & \textbf{33\%}         & 0.0002          & 8987      & \textbf{33\%}      & 0.0002       & 10136         & \textbf{18\%}         & 0.0002         \\
\tgt{Lsquic}                       & 5238             & 3294          & \textbf{59\%}         & 0.0002          & 3302      & \textbf{59\%}      & 0.0002       & 3759          & \textbf{39\%}         & 0.0002         \\
\tgt{Ngtcp2}                       & 3744             & 2882          & \textbf{30\%}         & 0.0002          & 3034      & \textbf{23\%}      & 0.0002       & 3570          & \textbf{5\%}          & 0.0006         \\ 
\tgt{Picoquic}                     & 4940             & 3329          & \textbf{48\%}         & 0.0002          & 3450      & \textbf{43\%}      & 0.0002       & 4014          & 23\%                  & \textcolor[rgb]{0.7,0.7,0.7}{0.1404}         \\ 
\tgt{Quicly}                       & 3291             & 2807          & \textbf{17\%}         & 0.0002          & 2808      & \textbf{17\%}      & 0.0002       & 2985          & \textbf{10\%}         & 0.0002         \\ 
\tgt{Xquic}                        & 6262             & 3569          & \textbf{75\%}         & 0.0002          & 3593      & \textbf{74\%}      & 0.0002       & 4205          & \textbf{49\%}         & 0.0002      \\ \cline{1-11}  
\end{tabular}%
}
\end{table*}

\subsection{Bug Case Studies}\label{apd:case-studies}

\noindent\Circled{M1} \textbf{Double-free error in Picotls when processing an unexpected public key length.}~
When Picotls encounters an unexpected value in the \texttt{x25519} public key length field of the Client Hello TLS message, it frees the memory allocated for the public key twice. Interestingly, the \tgt{Quicly} server also uses Picotls as its TLS library, but this bug was not observed in \tgt{Quicly} because we configured it to use the \texttt{secp256r1} elliptic curve group, which is unaffected by the issue. However, if we configure the \tgt{Quicly} server to use the \texttt{x25519} elliptic curve group, we can crash the server using the same crashing seed found on \tgt{Picoquic}. This vulnerability can therefore affect the \tgt{Picoquic} and \tgt{Quicly} servers equally depending on their TLS configuration.
\vspace{1mm}

\noindent\textit{\Circled{Impact}~}A remote attacker can exploit this vulnerability to conduct a DoS attack on a QUIC server using Picotls as the TLS library. In addition, according to~\cite{owasp_double_free}, freeing memory twice may also alter the program execution flow by overwriting particular registers or memory spaces, leading to arbitrary code execution. This vulnerability has been assigned a CVE ID (\textsf{CVE-2024-45402}) with a high severity score.

\vspace{3mm}
\noindent\Circled{M5} \textbf{Assertion failure in \tgt{Quicly} when processing an ACK frame in the draining state.}~ After the \tgt{Quicly} server receives a CONNECTION\_CLOSE frame, it enters a draining state, a state that ensures the remaining in-flight packets are processed before the connection is discarded. During this phase, the server will register a new callback function that will trigger an assertion failure when an acknowledgement is received. 

 The fuzzer exploits this vulnerability by sending two inputs in a sequence. First, the fuzzer sends an Initial packet containing a CRYPTO frame to initiate a handshake and a CONNECTION\_CLOSE frame to close the connection. The CRYPTO frame forces the server to respond with the necessary handshake messages, while the CONNECTION\_CLOSE puts the server into a draining state. Next, the fuzzer sends an Initial packet carrying an ACK frame to acknowledge the server's handshake messages. When the server attempts to process the acknowledgement, it invokes the previously mentioned callback function, precipitating an assertion failure.

\vspace{1mm}
\noindent\textit{\Circled{Impact}~}This vulnerability allows a remote attacker to trigger an assertion failure that crashes the \tgt{Quicly} server. This vulnerability has been assigned a CVE ID (\textsf{CVE-2024-45396}) with a high severity score.

\vspace{3mm}
\noindent\Circled{M6} \textbf{NULL pointer dereference in \tgt{Xquic} when handling a Stateless Reset packet on a destroyed connection.}~\tgt{Xquic} server (commit version ae6f7f7) crashes when it receives a Stateless Reset packet from the fuzzer. The packet carries a Stateless Reset Token tied to the Connection ID of the recently closed connection between the fuzzer and the server. Upon receiving the packet, the server checks the Stateless Reset Token against all connections, including recently closed connections. When it finds a match with the closed connection, it attempts to access the network path via a pointer that has already been freed and assigned a \texttt{NULL} value, leading to a NULL pointer dereference.

\vspace{1mm}
\noindent\textit{\Circled{Impact}~}A malicious actor can perform a DoS attack on the server by sending a Stateless Reset packet carrying a Stateless Reset Token associated with the Connection ID of a destroyed connection.

\vspace{3mm}
\noindent\Circled{M10} \textbf{NULL pointer dereference in \tgt{Xquic} when logging a Version Negotiation event.}~When a fuzzer sends an Initial packet with an incorrect QUIC version to trigger a Version Negotiation packet from the \tgt{Xquic} server (commit version 6803065), the server crashes. This is because a NULL pointer dereference occurs when the server attempts to log a Version Negotiation event. Specifically, the developers carelessly passed a \texttt{NULL} value as the network path pointer argument when calling the logging function (\texttt{xqc\_log\_event()}). 

\vspace{1mm}
\noindent\textit{\Circled{Impact}~}This allows an attacker to perform a DoS attack by sending an Initial packet with an incorrect QUIC version to the server.

\subsection{Ablation Study Discussion}\label{apd:ablation}

We report the ablation study results in Table~\ref{tab:quic-aflnet_ablation_study} in Section~\ref{sec:quic-aflent_stability}. We can observe each technique applied in \nameQUICAFLNet{} is effective and contributes to improving performance in a manner that is statistically significant. We discuss the results and observations further here.

\vspace{2mm}
\noindent\textbf{+QUIC-specific Cryptographic Module---addressing \Circled{C1}.~}The crypto module for handling QUIC-specific security functions described in Section~\ref{sec:quicCryptoModule} leads to statistically significant improvements across all targets. As a result of the addition, \nameQUICAFLNet{} achieves an increase of up to 62\% increase in branch coverage compared to the \aflnet{} (baseline). The crypto module we developed for QUIC aids the fuzzer in generating inputs with the correct integrity while ensuring the interpretability of decrypted data. This allows the fuzzer to reliably re-use the same test case to make incremental progress. This increases the likelihood of the fuzzer reaching deeper functionality in the target.

\vspace{2mm}
\noindent\textbf{+Synchronisation Protocol---addressing \Circled{C2}.~}Next, we activate the Synchronisation Protocol in addition to the previous configuration. As discussed in Section~\ref{sec:syncAndPerformance}, this module incorporates techniques to ensure that the fuzzing events are synchronised without manual interventions for timeout settings while mitigating delays impacting testing efficiency, which would otherwise reduce the fuzzer's throughput. Overall, the addition of the synchronisation protocol is seen to be effective at significantly improving median coverage metrics above those possible by adding cryptographic support. 

Interestingly, \tgt{Ngtcp2} does not show a statistically significant improvement in the median coverage metric. Although, in terms of the best coverage metrics for each configuration in the 10 trial, the +Synchronisation Protocol achieves higher coverage than the +QUIC-specific Cryptographic Module (3835 vs 3823), both configurations achieve similar median code coverage. In the previous configuration (+QUIC-specific Cryptographic Module), the manual timeout is empirically determined during a dry run phase for use during the fuzz testing phase. In the case of \tgt{Ngtcp2} implementation, the server implementation leads to more consistent behaviour in terms of processing various inputs; the empirically determined minimum timeout set to synchronise the fuzzing events in the dry run phase appears nearly to be optimal in this case. This consistency allows the fuzzer to re-use the same test case to reach its target state and explore further.

Overall, these results highlight that the performance impact which can arise from non-determinism, as discussed in \Circled{C3}, is mitigated by the synchronisation protocol. 

\vspace{2mm}
\noindent\textbf{+Snapshot Protocol---addressing \Circled{C3}.~}This configuration represents the full version of \nameQUICAFLNet{}. With the addition of the Snapshot protocol discussed in Section~\ref{sec:quicAflNetDesign}, the fuzzer mitigates the target initialisation overhead by spawning an initialised target instance from the snapshot captured during the initialisation of the Forkserver. Consequently, we can observe improvements in performance for targets that require long initialisation, such as \tgt{Lsquic}, where the difference is statistically significant.

\end{document}